\documentclass[11pt,english]{scrartcl}
\pdfoutput=1
\usepackage{ae,aecompl}
\usepackage{eulervm}
\usepackage[T1]{fontenc}
\usepackage[latin9]{inputenc}
\usepackage{amsmath}
\usepackage{amssymb}
\usepackage{graphicx}
\usepackage{wasysym}
\usepackage{esint}
\usepackage[authoryear]{natbib}

\usepackage{aecompl}
\usepackage{algorithm}
\usepackage{algpseudocode}

\usepackage{hyperref}
\hypersetup{
 colorlinks,linkcolor={black},citecolor={black},urlcolor={black},pdfencoding=auto}

\usepackage{babel}

\usepackage{babel}
\begin{document}
\renewcommand*{\thefootnote}{\fnsymbol{footnote}}
\title{Hydraulic Fracturing of Poorly Consolidated Reservoir During Waterflooding}
\author{Emmanuel Detournay\footnote{Corresponding author. Email address: detou001@umn.edu} \ and Yera Hakobyan}
\date{\small\textit{Department of Civil, Environmental, and Geo- Engineering, University of Minnesota, USA}}

\maketitle
\renewcommand*{\thefootnote}{\arabic{footnote}}

\begin{abstract}
This paper describes a KGD-type model of a hydraulic fracture created
by injecting fluid in weak, poorly consolidated rocks. The model is
based on the key assumption that the hydraulic fracture is propagating
within a domain where the hydraulic fields are quasi-stationary. By
further assuming a negligible toughness, the fracture is shown to
grow as a square root of time. The asymptotic fracture propagation
regimes at small and large time are constructed and the transient
solution is computed by solving a nonlinear system of algebraic equations
formulated in terms of the fracture aperture. At early time the fracture
is hydraulically invisible and the injection pressure increases with
time $t$ as $\log t$, while at late time leak-off from the borehole
is negligible and the injection pressure decreases as $t^{-1/4}$.
According to this model, the peak injection pressure observed when
injecting fluid in weak, poorly consolidated rocks should not be interpreted
as indicating a breakdown of the formation, but rather as marking
the transition between two asymptotic flow regimes. The timescale
that legislates the transition between the small and large time asymptotic
regimes is shown to be a strongly nonlinear function of a dimensionless
injection rate.
\end{abstract}

\section{Introduction}

The efficiency of waterflooding carried out to increase oil recovery
from hydrocarbon-bearing rocks is predicated in part on the initiation
and propagation of hydraulic fractures at injection wells to ensure
a more efficient sweep of the reservoir \citep{Sharma2000,VandenHoek2000,Noirot2003}.
These fracture allow the injected water to leak through the crack
surfaces, which eventually leads to the establisment of a quasi-linear
flow pattern around the borehole-fracture system. 

In the waterflooding of poorly consolidated reservoirs, unusually
large breakdown pressures are observed. This is a priori unexpected
in view of the small fracture toughness of these rocks. It has been
proposed that the observed anomalous breakdown pressure is a consequence
of the large apparent toughness associated with the development of
plastic zones in the crack tip region \citep{Cleary2007,Papanastasiou1993,Papanastasiou1997,VanDam2002}.
However, laboratory fluid injection experiments in weak sandstone
show evidence of injection-induced hairline cracks \citep{Ispas2012},
which contradicts the blunt crack tip predicted by plasticity-based
models \citep{Germanovich2012}. In view of this inconsistency, this
paper explores instead an alternative explanation, which is rooted
in the hydraulic nature of the problem rather than in the apparent
strength of the formation. 

Hydraulic fractures created during waterflooding of poorly consolidated
oil reservoirs cannot be analyzed, however, using models developed
for predicting the growth of hydraulic fractures in conventional reservoirs.
The reason is twofold. First, the volume balance equation of hydraulic
fracturing--- the injected fluid volume equals the crack volume plus
the leak-off volume --- is essentially meaningless. Indeed, the volume
of fluid stored in the crack is negligibly small compared to the volume
of injected fluid, in view of the large permeability of these reservoirs.
In other words the treatment efficiency --- the ratio of the stored
fluid volume to the injected volume --- is virtually zero. Second,
the conventional models do not account for the large scale perturbations
of the pore pressure due to the prolonged duration, of order of months
or years, of water injection.

The main objective of this paper is thus to develop a model for analyzing
the propagation of a KGD-type hydraulic fracture in the context of
a waterflooding operation. The paper is structured as follows. First
the problem is formulated within the framework of porous media flow,
lubrication theory, and linear elastic fracture mechanics. The model
is then expressed as a nonlinear system of integro-differential equations
in terms of the fracture aperture, by taking advantage of the linearity
of the equations governing the hydraulic and mechanical fields in
the porous medium. A scaling analysis indicates that the scaled fields
depend only on the dimensionless space and time variables, with all
the parameters of the problem absorbed in the scales. Small and large-time
asymptotic solutions are then constructed and the transient solution
is obtained by solving numerically a nonlinear system of algebraic
equations deduced by discretizing the integro-differential system
of equations governing the fracture aperture.

\section{Problem Definition}

\begin{figure}[h]
\begin{centering}
\includegraphics[scale=0.7]{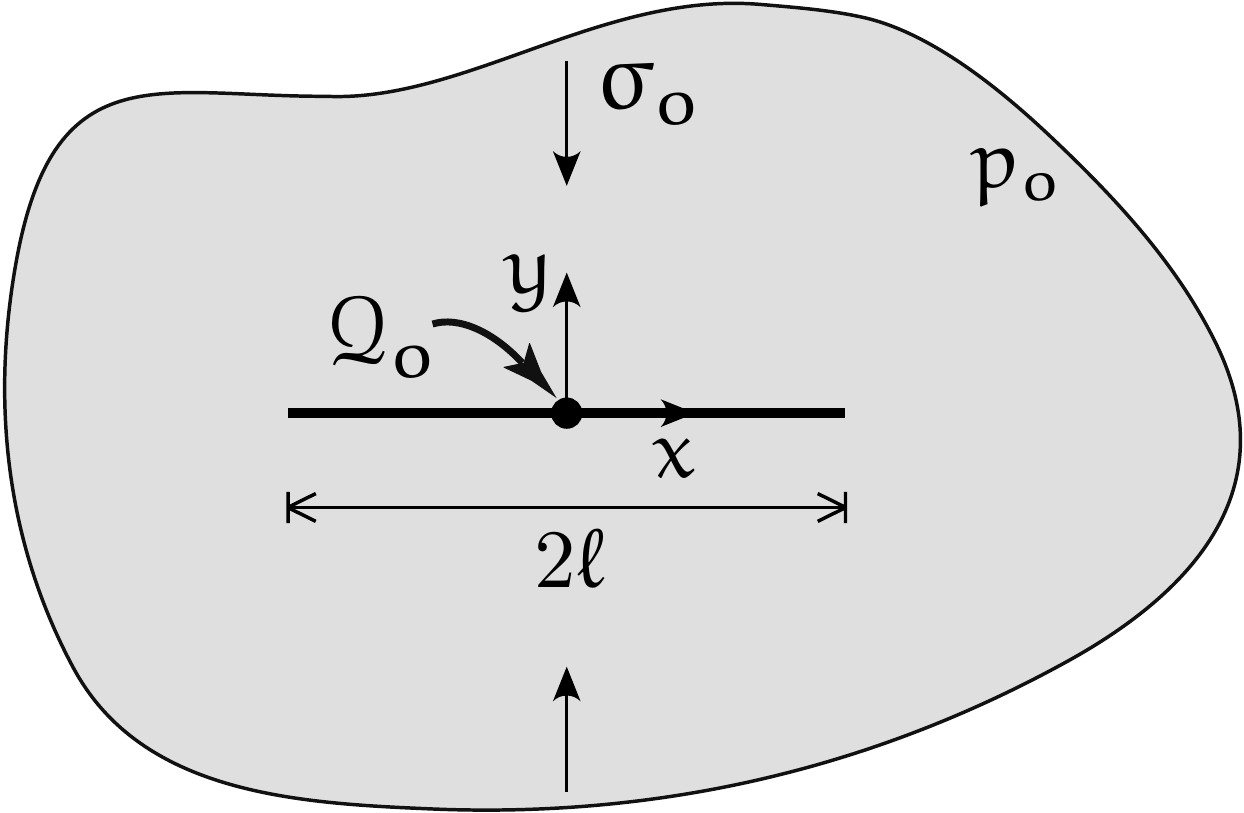} 
\par\end{centering}
\caption{Problem definition. The model is constructed assuming that the borehole
can be reduced to a point source. }
\label{fig1} 
\end{figure}

We consider the propagation of a KGD-type (plane strain) hydraulic
fracture driven by injection of a Newtonian fluid in a permeable elastic
rock, as sketched in Fig. \ref{fig1}. The far-field boundary conditions
correspond to minimum compressive stress $\sigma_{o}$ (normal to
the hydraulic fracture) and pore pressure $p_{o}$. A fluid of viscosity
$\mu$ is injected at a constant rate $\mathcal{Q}_{o}$ per unit
length of the hole. Prior to injection, the stress and pore pressure
fields are uniform. The rock is characterized by intrinsic permeability
$k$, or equivalently by mobility $\kappa=k/\mu$, diffusivity $c$,
elastic modulus $E$, and Poisson ratio $\nu$. The rock toughness
is assumed to be negligible. 

In view of the plane strain nature of the problem, the two elastic
constants $E$ and $\nu$ can be combined into the plane strain modulus
$E'$. Furthermore, to avoid carrying numerical factors in the equations,
we introduce the alternate constant $\mu'$. The constants $E'$ and
$\mu'$ are defined as 
\begin{equation}
E'=\frac{E}{1-\nu^{2}}\qquad\mu^{\prime}=12\mu\label{mf1}
\end{equation}
The final set of model parameters consists therefore of 7 constants:
$\mathcal{P}=\{\mu'$, $E'$, $\kappa$, $c$, $\sigma_{o}$, $p_{o}$,
$\mathcal{Q}_{o}\}$. Using Buckingham $\Pi$-theorem, this set can
in principle be reduce to 4 dimensionless parameters. However, further
analysis of the problem indicates that a much greater simplification
of the problem is possible.

The hydraulic fracturing problem can be formulated as follows. Given
the fluid and solid properties, the far-field stress and pore pressure,
and the injection rate, determine how the crack half-length (one wing)
$\ell(t)$ and injection pressure $p_{w}(t)$ evolve with time $t$.

\section{Mathematical Model}

Here we formulate a mathematical model that combines (i) the diffusion
equation controlling the pore pressure evolution in the permeable
medium; (ii) the equations of Linear Elastic Fracture Mechanics (LEFM),
one linking crack aperture and fluid pressure, the other defining
the propagation criterion; (iii) the lubrication equation governing
fluid flow in the fracture; and finally (iv) the boundary conditions
and interface conditions between the fracture and the porous medium. 

The model is based on the following assumptions: (i) the fracture
length is large compared to the borehole radius so that the fluid
is injected via a point source; (iii) the viscosity of the injected
fluid is identical to that of the pore fluid; (iv) the crack propagates
inside the region where the flow is quasi-stationary (i.e., where
the pore pressure is governed by Laplace equation); (v) the crack
volume is negligible compared to the volume of fluid injected, or
equivalently, the rate of change of the fluid volume stored in the
crack is negligible compared to the injection rate.

Consider the system of coordinates $(x,y)$ with the origin at the
point source and the $x$-axis oriented in the direction normal to
the minimum in-situ stress $\sigma_{o}$. Hence, as illustrated in
Fig. \ref{fig1}, the fracture propagates along the $x$-axis. Pore
pressure $p_{r}(x,y,t)$ is defined in the plane $(x,y)$, while fluid
pressure $p_{f}(x,t)$ or equivalently net pressure $p(x,t)=p_{f}(x,t)-\sigma_{o}$,
crack aperture $w(x,t)$, and leak-off $g(x,t)$ are defined along
the crack $-\ell(t)\le x\le\ell(t)$, $y=0$. 

\subsection{Diffusion}

The pore pressure field $p_{r}(x,y,t)$ in the permeable medium is
governed by the diffusion equation 
\begin{equation}
c\nabla^{2}p_{r}-\frac{\partial p_{r}}{\partial t}=\frac{1}{S}\left[g(x,t)\delta(y)+\mathcal{Q}_{o}(1-\Phi(t))\delta(x,y)\right],\label{eq:mf2}
\end{equation}
where $\delta(y)$ and $\delta(x,y)$ denote the Dirac $\delta$-function
in one- and two-dimensional space, respectively. The storage coefficient
is defined as $S=\kappa/c$. The leak-off $g(x,t)$ represents the
normal flux discontinuity across the crack and corresponds therefore
to a source density ($g>0$) from the point of view of the porous
medium.

The presence of the crack causes the injection of fluid into the rock
to be partly distributed along the crack rather than to take place
solely at the source. The unknown function $\Phi(t)$ characterizes
the partitioning of the injected fluid between the crack and the porous
medium at the injection point. Thus, $0\le\Phi(t)\le1$ with the lower
limit corresponding to a fracture of relatively zero conductivity
and the upper limit to a fracture of relatively infinite conductivity.
Also, $\Phi(t)=0$ for $\tau\le0$ as injection starts at $t=0$.
Moreover, because storage of fluid in the fracture is negligible on
the ground of the high permeability of the porous medium, the leak-off
$g(x,t)$ is constrained by 
\begin{equation}
\int_{-\ell(t)}^{\ell(t)}g(x,t)\mathrm{d}x=\Phi(t)\mathcal{Q}_{o}.\label{eq:mf12}
\end{equation}

The pore pressure field $p_{r}(x,y,t)$ is subject to the conditions
at infinity 
\begin{equation}
\lim_{r\rightarrow\infty}p_{r}=p_{o},\label{eq:mf3}
\end{equation}
where $r=\sqrt{x^{2}+y^{2}}$ and to the initial conditions, here
assumed to correspond to a uniform pressure field $p_{o}$ 
\begin{equation}
p_{r}(x,y,0)=p_{o}.\label{eq:mf4}
\end{equation}

\subsection{Lubrication}

The flow of fluid in the crack is governed by Reynolds lubrication
equation \citep{Batc67}
\begin{equation}
\frac{1}{\mu'}\frac{\partial}{\partial x}\left(w^{3}\frac{\partial p_{f}}{\partial x}\right)=g(x,t)-\mathcal{Q}_{o}\Phi(t)\delta(x),\label{eq:mf5}
\end{equation}
which is obtained by combining Poiseuille's law 
\begin{equation}
q=-\frac{w^{3}}{\mu'}\frac{\partial p_{f}}{\partial x}\label{eq:mf6}
\end{equation}
with the continuity equation 
\begin{equation}
\frac{\partial q}{\partial x}+g-\mathcal{Q}_{o}\Phi(t)\delta(x)=0.\label{eq:mf7}
\end{equation}
The source term in the continuity equation implies a flux discontinuity
at the origin, which can be translated as 
\begin{equation}
q(0^{\pm},t)=\pm\frac{1}{2}\mathcal{Q}_{o}\Phi(t)\label{eq:mf9}
\end{equation}
in view of the problem symmetry. Finally, the flux $q(x,t)$ in the
crack vanishes at the crack tip \citep{DePe14}
\begin{equation}
q(\pm\ell,t)=0.\label{eq:mf8}
\end{equation}

The fluid pressure $p_{f}$ in the crack is continuous with the pore
pressure field 
\begin{equation}
p_{f}(x,t)=p_{r}(x,0,t)\quad|x|\le\ell(t).\label{eq:mf13}
\end{equation}
This equation together with the continuity of the normal flux at the
crack walls, which is implicitly satisfied by equating source density
along the crack to the leak-off rate, as done in \eqref{eq:mf2},
represent the two hydraulic conditions at the interface between the
crack and the porous medium.

\subsection{Linear Elastic Fracture Mechanics}

The net pressure in the crack and the crack aperture are also related
by a singular integral equation obtained by superposition of dislocation
dipoles
\begin{equation}
p_{f}(x,t)-\sigma_{o}=-\frac{E'}{4\pi}\int_{-\ell(t)}^{\ell(t)}\frac{w(s,t)}{(x-s)^{2}}\mathrm{d}s.\label{eq:mf14}
\end{equation}
According to the theory of singular integral equations \citep{Musk46},
the above equation necessarily implies that the crack closes at its
tip, i.e., $w(\pm\ell(t),t)=0$. Finally, the propagation criterion
$K_{I}=0$ can be expressed as a weighted integral of the net pressure
\eqref{eq:mf14} \citep{Buec70,Rice72}

\begin{equation}
\int_{0}^{\ell(t)}\frac{\left(p_{f}(x,t)-\sigma_{o}\right)\mathrm{d}x}{\sqrt{\ell^{2}(t)-x^{2}}}=0.\label{eq:mf15}
\end{equation}

To simplify the writing of equations, we introduce the net pressure
$p(x,t)$ defined as 
\begin{equation}
p=p_{f}-\sigma_{o}.\label{eq:mf17}
\end{equation}

\section{Asymptotic Mathematical Model}

The problem can thus be formulated only in terms of variables that
are defined on the 1D crack. The system of equations \eqref{eq:mf2}-\eqref{eq:mf5},
\eqref{eq:mf8}-\eqref{eq:mf9}, and \eqref{eq:mf14}-\eqref{eq:mf15},
is closed and can be used to evolve the solution, in particular crack-wing
length $\ell(t)$, and fields $p(x,t)$, $w(x,t)$, and $g(x,t)$.
The formulation can be simplified, however, by further adopting the
assumption that the crack propagates in a region where the pore pressure
is quasi-stationary. As a result, the problem loses its evolutionary
nature: time is no longer an independent variable, but becomes a parameter
of the problem. This change in the nature of the problem will be reflected
in the notation of the fields defined along the crack by using a semi-colon
rather than a coma in front of $t$, i.e., $p(x;t)$, $w(x;t)$, and
$g(x;t)$.

\subsection{Decomposition of the pore pressure field}

First, we express the pore pressure field $p_{r}(x,y,t)$ as the superposition
of three fields 
\begin{equation}
p_{r}(x,y,t)=p_{o}+\Delta p_{w}(r,t)+\Delta p_{c}(x,y,t)\label{eq:am1}
\end{equation}
where $\Delta p_{w}(r,t)$ is the pore pressure field induced by injection
of fluid in the absence of a crack, $\Delta p_{c}(x,y,t)$ is the
pore pressure perturbation caused by the exchange of fluid between
the crack and the medium, and $p_{o}$ is the initial homogeneous
field at the onset of injection. As a consequence of the linearity
of the diffusion equation, the two fields $\Delta p_{w}$ and $\Delta p_{c}$
are respectively governed by 
\begin{equation}
c\nabla^{2}\Delta p_{w}-\frac{\partial\Delta p_{w}}{\partial t}=\frac{\mathcal{Q}_{o}}{S}H(t)\delta(x,y)\label{eq:am2}
\end{equation}
and 
\begin{equation}
c\nabla^{2}\Delta p_{c}-\frac{\partial\Delta p_{c}}{\partial t}=\frac{1}{S}\left[g(x,t)\delta(y)-\mathcal{Q}_{o}\Phi(t)\delta(x,y)\right]\label{eq:am3}
\end{equation}
noting also the initial conditions 
\begin{equation}
\Delta p_{w}=0\qquad\Delta p_{c}=0,\quad t=0\label{eq:am4}
\end{equation}
and the boundary conditions at infinity 
\begin{equation}
\Delta p_{w}=0\qquad\Delta p_{c}=0,\quad r\rightarrow\infty\label{eq:am5}
\end{equation}
The magnitude of the singular source terms in \eqref{eq:am2} and
\eqref{eq:am3} is consistent with our definition of $\Delta p_{w}$
and $\Delta p_{c}$; i.e., $\Delta p_{w}$ representing the injection-induced
pore pressure in the absence of a crack, and $\Delta p_{c}$ representing
the additional pore pressure perturbation associated with the existence
of a fracture. It will be shown that the far-field signature of the
source terms in \eqref{eq:am3} ---the right-hand member of \eqref{eq:am3}---
is a singular double dipole at the origin.

The pore pressure field in the absence of a crack is given by the
well-known source solution, often referred to as Thys solution in
geo-hydrology \citep{ChDe98} 
\begin{equation}
\Delta p_{w}=\frac{\mathcal{Q}_{o}}{4\pi\kappa}E_{1}\left(\frac{r^{2}}{4ct}\right)\label{eq:am6}
\end{equation}
where $E_{1}$ is the exponential integral. This solution has the
interesting property, deduced from the asymptotics of $E_{1}$, that
inside a radius $r_{s}=\chi_{s}\sqrt{ct}$ with $\chi_{s}$ of order
$O(10^{-1}),$ $\Delta p_{w}$ can be approximated as 
\begin{equation}
\Delta p_{w}\simeq-\frac{\mathcal{Q}_{o}}{2\pi\kappa}\left(\ln\frac{r}{2\sqrt{ct}}+\frac{\gamma}{2}\right),\qquad r<r_{s}\label{eq:am7}
\end{equation}
In the above, $\gamma=0.5772...$ is Euler's constant. With an accuracy
of about 1\%, $\Delta p_{w}$ can be approximated by the asymptotic
solution inside $r=r_{s}$ if $\chi_{s}\approx0.35.$

Thus inside the quasi-stationary region $r<r_{s}(t)$, the pore pressure
evolves only because of the movement of the diffusion front. Provided
that the crack is well inside the quasi-stationary region, we can
assume that there are no transients associated with the exchange of
fluid between the crack and the medium. Thus, we assume that 
\begin{equation}
\nabla^{2}\Delta p_{c}\simeq\frac{1}{\kappa}\left[g(x,t)\delta(y)-\mathcal{Q}_{o}\Phi(t)\delta(x,y)\right],\qquad r<r_{s}(t)\label{eq:am9}
\end{equation}
noting that the leak-off function $g(x,t)$ is only defined along
the crack. Hence the pore pressure field $\Delta p_{c}$ can be written
as an integral expression using a superposition of source solution
of the Laplace equation, i.e., 
\begin{equation}
\Delta p_{c}(x,y;t)=\int_{-\ell(t)}^{\ell(t)}P_{s}(x-s,y)g(s;t)\mathrm{d}s-\mathcal{Q}_{o}\Phi(t)P_{s}(x,y)\label{eq:am10}
\end{equation}
where the source influence function $P_{s}$ is given by 
\begin{equation}
P_{s}(x,y)=-\frac{1}{2\pi\kappa}\ln r,\qquad r=\sqrt{x^{2}+y^{2}}.\label{eq:am11}
\end{equation}
It is important to stress that time only enters the expression for
$\Delta p_{c}$ via the bounds of the integral and through the dependence
of the leak-off on time. Pore pressure component $\Delta p_{c}$ decays
as $1/r^{2}$ away from the crack, as the far-field hydraulic effect
of the crack is equivalent to a double dipole.

\subsection{Reformulation}

It is now possible to reformulate the problem only in terms of the
net pressure $p(x;t)$, crack opening $w(x;t)$, leak-off $g(x;t)$,
fluid fraction $\Phi(t)$, and crack length $\ell(t)$. Thanks to
the integral representation of $p(x;t)$ in terms of leak-off $g(x;t)$
and crack aperture $w(x;t)$, we can reduce the formulation of the
problem to a set of integro-differential equations on the crack and
to unknowns only associated with the crack. With this formulation,
the elastic fields and pore pressure field in the domain containing
the crack are calculated by solving singular integrals equations on
the crack. These integral equations make use of free-field singular
solutions: dislocation dipoles for elasticity and source solutions
for porous media flow.

The set of equations is given by

\begin{equation}
\frac{1}{\mu'}\frac{\partial}{\partial x}\left(w^{3}\frac{\partial p}{\partial x}\right)=g(x;t)-\mathcal{Q}_{o}\Phi(t)\delta(x),\label{eq:re1}
\end{equation}
\begin{equation}
\mathcal{Q}_{o}\Phi(t)=\int_{-\ell(t)}^{\ell(t)}g(x;t)\mathrm{d}x\label{eq:re8}
\end{equation}

\begin{equation}
p(x;t)+\sigma_{o}-p_{o}-\Delta\bar{p}_{w}(x;t)=-\frac{1}{2\pi\kappa}\int_{-\ell(t)}^{\ell(t)}\ln|x-s|\,g(s;t)\mathrm{d}s+\frac{\mathcal{Q}_{o}}{2\pi\kappa}\Phi(t)\ln|x|,\label{eq:re2}
\end{equation}

\begin{equation}
p(x;t)=-\frac{E'}{4\pi}\int_{-\ell(t)}^{\ell(t)}\frac{w(s;t)}{(x-s)^{2}}\mathrm{d}s,\label{eq:re3}
\end{equation}
where $\Delta\bar{p}_{w}(x,t)=\Delta p_{w}(r,t)|_{y=0}$ reads 
\begin{equation}
\Delta\bar{p}_{w}(x,t)=-\frac{\mathcal{Q}_{o}}{2\pi\kappa}\left(\ln\frac{|x|}{2\sqrt{ct}}+\frac{\gamma}{2}\right).\label{eq:re5}
\end{equation}
The propagation criterion is expressed in terms of a weighted integral
of the net pressure 
\begin{equation}
\int_{0}^{\ell(t)}\frac{p(x;t)\mathrm{d}x}{\sqrt{\ell^{2}(t)-x^{2}}}=0,\label{eq:re4}
\end{equation}
Note that \eqref{eq:re8} implies that the flux vanishes at the crack
tip 
\begin{equation}
w^{3}\frac{\partial p}{\partial x}=0,\quad x=\pm\ell(t),\label{eq:re6}
\end{equation}
as can readily be verified by integrating \eqref{eq:re1} using \eqref{eq:re8}.
The condition $w(\pm\ell,t)=0$ is also automatically met by the integral
equation \eqref{eq:re3}.

The above system of equations \eqref{eq:re1}-\eqref{eq:re6} is closed.
Hence, given $t$ and the set of parameters $\mathcal{P}=\{\mu',\kappa,c,E',\sigma_{o},p_{o},\mathcal{Q}_{o}\}$,
the solution $\ \mathcal{S}(x;t)=\{w(x;t)$, $p(x;t)$, $\,g(x;t)$,
$\Phi(t)$, $\ell(t)\}$ can be determined from the system of equations
\eqref{eq:re1}-\eqref{eq:re6}. In other words, $\mathcal{S}(x;t)$
can be calculated without the need to know the solution at prior times.

\section{Scaling}

Scaling indicates that this problem is characterized by one time scale,
$t_{K}$, which is associated with the transition between the rock-flow
to the fracture-flow regime.

\subsection{Scales}

First we introduce scales for the aperture, pressure, leak-off, length,
and time. These scales, respectively denoted as $w_{K}$, $p_{K}$,
$g_{K}$, $\ell_{K}$, $t_{K}$, are undetermined for the time being.
Dimensionless crack aperture $\Omega(\xi;\tau)$, net pressure $\Pi(\xi;\tau)$,
flux $\Psi(\xi;\tau)$, leak-off rate $\Gamma(\xi;\tau)$, and crack
length $\Lambda(\tau)$ are then naturally defined as 
\begin{equation}
\Omega=\frac{w}{w_{K}},\qquad\Pi=\frac{p}{p_{K}},\qquad\Psi=\frac{q}{Q_{0}},\qquad\Gamma=\frac{g}{g_{K}},\qquad\Lambda=\frac{\ell}{\ell_{K}\tau^{1/2}}\label{eq:sc1}
\end{equation}
with scaled coordinate $\xi$ along the crack and dimensionless time
$\tau$ given by 
\begin{equation}
\xi=\frac{x}{\ell(t)},\qquad\tau=\frac{t}{t_{k}}\label{eq:sc2}
\end{equation}

The motivation to define $\Lambda$ according to \eqref{eq:sc1} stems
from the search of similarity solutions at small and large time. Indeed,
the existence of similarity solutions requires that $\ell\sim t^{1/2}$.
We will show that $\Lambda=1$ if the toughness is zero.

\subsection{Dimensionless Groups}

The system of equations \eqref{eq:re1}-\eqref{eq:re5} is rewritten
in terms of the newly introduced quantities

\begin{equation}
\frac{\mathcal{G}_{m}}{\tau\Lambda^{2}}\frac{\partial}{\partial\xi}\left(\Omega^{3}\frac{\partial\Pi}{\partial\xi}\right)=\Gamma-\frac{\mathcal{I}\mathcal{G}_{i}}{\mathcal{G}_{c}}\frac{\Phi\delta(\xi)}{\tau^{1/2}\Lambda}\label{eq:sc4}
\end{equation}
\begin{equation}
\frac{\mathcal{I}\mathcal{G}_{i}}{\mathcal{G}_{c}}\Phi(\tau)=\tau^{1/2}\Lambda\int_{-1}^{1}\Gamma(\zeta)\mathrm{d}\zeta\label{eq:sc20}
\end{equation}

\begin{equation}
\Pi-\Pi_{o}=-\frac{\mathcal{G}_{c}\tau^{1/2}\Lambda}{2\pi}\int_{-1}^{1}\ln|\xi-\zeta|\,\Gamma(\zeta)\mathrm{d}\zeta+\frac{\mathcal{I}\mathcal{G}_{i}}{2\pi}\Phi\ln|\xi|\label{eq:sc5}
\end{equation}
\begin{equation}
\Pi=-\frac{\mathcal{G}_{e}}{4\pi\tau^{1/2}\Lambda}\int_{-1}^{1}\frac{\Omega\mathrm{d}\zeta}{\left(\xi-\zeta\right)^{2}}\label{eq:sc6}
\end{equation}
\begin{equation}
\int\frac{\Pi\mathrm{d}\xi}{\sqrt{1-\xi^{2}}}=0\label{eq:sc7}
\end{equation}
where

\begin{equation}
\Pi_{o}=-\mathcal{G}_{i}\left[1+\frac{\mathcal{I}}{2\pi}\left(\ln\frac{\Lambda|\xi|}{2\mathcal{G}_{l}}+\frac{\gamma}{2}\right)\right]\label{eq:sc8}
\end{equation}
noting that $\delta(\xi)=\tau^{1/2}\Lambda\ell_{K}\delta(x)$. The
dimensionless injection rate $\mathcal{I}$ appearing in \eqref{eq:sc4}-\eqref{eq:sc8}
is defined as 

\begin{equation}
\mathcal{I}=\frac{\mathcal{Q}_{o}}{\kappa(\sigma_{o}-p_{o})}\label{eq:sc14}
\end{equation}
Thus, five dimensionless groups emerge from the above equations:

\begin{align}
\mathcal{G}_{m} & =\frac{w_{K}^{3}p_{K}}{\mu'g_{K}\ell_{K}^{2}},\quad\mathcal{G}_{i}=\frac{\sigma_{o}-p_{o}}{p_{K}},\quad\mathcal{G}_{c}=\frac{g_{K}\ell_{K}}{\kappa p_{K}},\quad\mathcal{G}_{e}=\frac{E'w_{K}}{\ell_{K}p_{K}},\quad\mathcal{G}_{l}=\frac{\sqrt{ct_{K}}}{\ell_{K}}\label{eq:sc9}
\end{align}

\subsection{Characteristic quantities}

The five scales $w_{k}$, $p_{k}$, $g_{K}$, $\ell_{K}$, and $t_{K}$
become explicit by imposing a value to the five dimensionless groups
defined in \eqref{eq:sc9}. While there is arbitrariness in the chosen
values for these groups, they should in principle be selected in such
a way that the norm of the fields $\Omega(\xi;\tau)$, $\Pi(\xi;\tau)$,
$\Gamma(\xi;\tau)$, as well as $\Lambda(\tau)$ are of order $O(1)$
when $\tau=1$.

Here we choose 
\begin{equation}
\mathcal{G}_{m}=1,\quad\mathcal{G}_{e}=1,\quad\mathcal{G}_{c}=1,\quad\mathcal{G}_{i}=\frac{1}{\mathcal{I}},\quad\mathcal{G}_{l}=\frac{1}{4}\exp\left(\frac{2\pi}{\mathcal{I}}+\frac{\gamma}{2}\right)\label{eq:sc10}
\end{equation}
As a result of enforcing \eqref{eq:sc10}, explicit expressions for
the scales are obtained: 
\begin{equation}
w_{K}=\frac{\ell_{K}\mathcal{Q}_{o}}{\kappa E'},\quad p_{K}=\frac{\mathcal{Q}_{o}}{\kappa},\quad g_{K}=\frac{\mathcal{Q}_{o}}{\ell_{K}},\quad\ell_{K}=\left(\frac{\mu'\kappa^{4}E'^{3}}{\mathcal{Q}_{o}^{3}}\right)^{1/2},\quad t_{K}=\frac{\mu'\kappa^{4}E'^{3}}{16c\mathcal{Q}_{o}^{3}}\exp\left(\frac{4\pi}{\mathcal{I}}+\gamma\right)\label{eq:sc13}
\end{equation}
Note that with this scaling, expression \eqref{eq:sc8} for $\Pi_{o}(\xi)$
reduced to
\begin{equation}
\Pi_{o}(\xi)=-\frac{1}{2\pi}\ln(2\Lambda|\xi|)\label{eq:sc15}
\end{equation}

\subsection{Scaled Mathematical Model}

It is noteworthy to point out that the scaled solution $\mathcal{S}(\xi;\tau)=\{\Omega,\,\Pi,\,\Gamma,\,\Phi,\,\Lambda\}$
depends only on the spatial variable $\xi$ and the time parameter
$\tau$, as all the physical parameters have been absorbed in the
scales. The solution is obtained by solving the closed system of integro-differential
equations 
\begin{equation}
\frac{1}{\tau^{1/2}\Lambda}\frac{\partial}{\partial\xi}\left(\Omega^{3}\frac{\partial\Pi}{\partial\xi}\right)=\tau^{1/2}\Lambda\Gamma-\Phi(\tau)\delta(\xi)\label{eq:sm1}
\end{equation}

\begin{equation}
\Phi(\tau)=\tau^{1/2}\Lambda\int_{-1}^{1}\Gamma(\zeta)\mathrm{d}\zeta\label{eq:sm5}
\end{equation}

\begin{equation}
\Pi(\xi)=-\frac{1}{2\pi}\left[\ln(2\Lambda)+(1-\Phi(\tau))\ln|\xi|+\tau^{1/2}\Lambda\int_{-1}^{1}\ln|\xi-\zeta|\,\Gamma(\zeta)\mathrm{d}\zeta\right]\label{eq:sm2}
\end{equation}
\begin{equation}
\Pi(\xi)=-\frac{1}{4\pi\tau^{1/2}\Lambda}\int_{-1}^{1}\frac{\Omega(\zeta)\mathrm{d}\zeta}{(\xi-\zeta)^{2}}\label{eq:sm3}
\end{equation}
\begin{equation}
\int_{0}^{1}\frac{\Pi\mathrm{d}\xi}{\sqrt{1-\xi^{2}}}=0\label{eq:sm4}
\end{equation}

Interestingly, the solution $\Pi(\xi)$ of the reduced system of equations
\eqref{eq:sm1}-\eqref{eq:sm3} automatically satisfies the propagation
criterion \eqref{eq:sm4} if $\Lambda=1$. This result can readily
be confirmed at small time, when $\Pi(\xi)\simeq\Pi_{o}(\xi)$ as
shown in Section \eqref{subsec:R-Regime}. Indeed, replacing $\Pi(\xi)$
in the propagation criterion \eqref{eq:sm4} by expression \eqref{eq:sc15}
implies that $\Lambda=1.$ Although this result has not be formally
proven to hold at any time, it has been comfirmed numerically. In
the following, we will impose that $\ell=\tau^{1/2}$ and thus remove
the propagation criterion \eqref{eq:sm4} from the set of governing
equations.

\subsection{Alternative Scaled Mathematical Model\label{subsec:Alternative-Scaled-Mathematical}}

The system of equations \eqref{eq:sm1}-\eqref{eq:sm4} can be rewritten
so as to remove $\Phi(\tau)$ from the formulation. To that effect,
express $\Gamma(\xi;\tau)$ as the superposition of two functions
\begin{equation}
\Gamma(\xi;\tau)=\tilde{\Gamma}(\xi;\tau)+\frac{1}{\tau^{1/2}}\Phi(\tau)\delta(\xi)\label{eq:sm11}
\end{equation}
Substituting \eqref{eq:sm11} into \eqref{eq:sm5} shows that 
\begin{equation}
\int_{-1}^{1}\tilde{\Gamma}(\zeta)\mathrm{d}\zeta=0\label{eq:sm18}
\end{equation}
Hence, the system of equations to be solved for $\tilde{\mathcal{S}}(\xi;\tau)=\{\Omega,\,\Pi,\,\tilde{\Gamma}\}$
is 
\begin{equation}
\frac{1}{\tau}\frac{\partial}{\partial\xi}\left(\Omega^{3}\frac{\partial\Pi}{\partial\xi}\right)=\tilde{\Gamma}\label{eq:sm12}
\end{equation}

\begin{equation}
\Pi(\xi)=\Pi_{o}(\xi)-\frac{1}{2\pi}\tau^{1/2}\int_{-1}^{1}\ln|\xi-\zeta|\,\tilde{\Gamma}(\zeta)\mathrm{d}\zeta\label{eq:sm13}
\end{equation}
\begin{equation}
\Pi(\xi)=-\frac{1}{4\pi\tau^{1/2}}\int_{-1}^{1}\frac{\Omega(\zeta)\mathrm{d}\zeta}{(\xi-\zeta)^{2}}\label{eq:sm14}
\end{equation}
\begin{equation}
\Omega^{3}\frac{\partial\Pi}{\partial\xi}=0,\quad\xi=\pm1\label{eq:sm19}
\end{equation}
where 
\begin{equation}
\Pi_{o}(\xi)=-\frac{1}{2\pi}\ln(2|\xi|)\label{eq:sm16}
\end{equation}
The constraint \eqref{eq:sm18} is automatically satisfied by the
zero-flux boundary condition \eqref{eq:sm19} at the crack tip and
Reynolds equation \eqref{eq:sm12}. 

\subsection{Intermediate-Range Pore Pressure Field}

The existence of a hydraulic fracture causes a perturbation of the
pore pressure field, $\Delta\Pi(\xi,\eta)=\Pi-\Pi_{o}$ given by 
\begin{equation}
\Delta\Pi=-\frac{1}{2\pi}\tau^{1/2}\int_{-1}^{1}\ln\sqrt{(\xi-\zeta)^{2}+\eta^{2}}\,\tilde{\Gamma}(\zeta)\mathrm{d}\zeta\label{eq:ip1}
\end{equation}
At distance large enough from the crack but still within the quasi-stationary
region, the pore pressure perturbation $\Delta\Pi$ is equivalent
the pore pressure induced by a steady-state quadripole centered at
the origin and aligned along the $\xi$-axis, i.e., 
\begin{equation}
\Delta\Pi\simeq\frac{1}{4\pi}\tau^{1/2}\mathcal{M}^{(2)}\frac{\xi^{2}-\eta^{2}}{\left(\xi^{2}+\eta^{2}\right)^{2}}\label{eq:ip2}
\end{equation}
where $\mathcal{M}^{(2)}$ is the second moment of leak-off $\tilde{\Gamma}(\xi)$
over the crack length 
\begin{equation}
\mathcal{M}^{(2)}=\int_{-1}^{1}\xi^{2}\tilde{\Gamma}(\xi)\,\mathrm{d}\xi\label{eq:ip3}
\end{equation}
Expression \eqref{eq:ip2} is an intermediate asymptotic solution.
It is worth noting that $\tilde{\Gamma}(\xi)$ and $\Gamma(\xi)$
have identical moments, except for the zeroth moment since the two
functions differ by a Dirac $\delta$-function. Indeed, $\tilde{\mathcal{M}}^{(0)}=0$
while $\mathcal{M}^{(0)}=\Phi(\tau)$.

\subsection{Discussion}

As shown above, the model can be formulated either in terms of $\Gamma(\xi)$
or $\tilde{\Gamma}(\xi)$. Consider first the model formulated in
terms of $\Gamma(\xi;\tau)$ and $\Phi(\tau)$. The porous medium
equation \eqref{eq:sm2} indicates that the pore pressure field induced
by injection of fluid is the superposition of a discrete continuous
source of strength $(1-\Phi(\tau))$ and of a distributed source density
$\Gamma(\xi;\tau)$ with a combined strength $\Phi(\tau)$ 
\begin{equation}
\tau^{1/2}\int_{-1}^{1}\Gamma(\zeta;\tau)\mathrm{d}\zeta=\Phi(\tau),\qquad\Psi(0^{\pm},\tau)=\pm\frac{1}{2}\Phi(\tau)\label{eq:sm20}
\end{equation}
The above equation expresses the property that all the fluid entering
the crack at the inlet leaves the fracture via leak-off.

Formulating the model in terms of $\tilde{\Gamma}(\xi;\tau)$ implies
instead that fluid enters the fracture via the leak-off function ---
as a Dirac, and is then expelled along the crack into the porous medium.
Hence, 
\begin{equation}
\int_{-1}^{1}\tilde{\Gamma}(\zeta;\tau)\mathrm{d}\zeta=0,\qquad\tilde{\Psi}(0^{\pm},\tau)=0\label{eq:sm21}
\end{equation}

The two approaches are essentially equivalent in regard to the determination
of the crack opening $\Omega(\xi;\tau)$ and net pressure $\Pi(\xi;\tau)$.
However, the injection pressure is directly affected by the strength
of the logarithmic singularity of $\Pi$, which depends directly on
$\Phi$, as can be surmised from \eqref{eq:sm2}. A correct evaluation
of the fluid fraction $\Phi$ requires, therefore, consideration of
the borehole radius $a$.

\section{Asymptotic Regimes}

\subsection{Similarity Solutions with Power Law Dependence on Time\label{subsec:Similarity-Solutions}}

For small- and large-time, one searches for solution of the type 
\begin{equation}
\Omega=\Omega_{*}(\xi)\tau^{\beta_{*}},\quad\Pi=\Pi_{*}(\xi)\tau^{\alpha_{*}},\quad\Gamma=\Gamma_{*}(\xi)\tau^{\gamma_{*}},\quad\Phi=\Phi_{*}\tau^{\delta_{*}}\label{eq:ar1}
\end{equation}
Note that the existence of a similarity solution hinges necessarily
on the requirement that the fracture length grows at the square root
or time, which is indeed met at all time for $K_{Ic}=0$. Using \eqref{eq:ar1},
the system of equations \eqref{eq:sm1}-\eqref{eq:sm4} becomes

\begin{equation}
\tau^{\alpha_{*}+3\beta_{*}-1/2}\frac{\mathrm{d}}{\mathrm{d}\xi}\left(\Omega_{*}^{3}\frac{\mathrm{d}\Pi_{*}}{\mathrm{d}\xi}\right)=\tau^{\gamma_{*}+1/2}\Gamma_{*}-\tau^{\delta_{*}}\Phi_{*}\delta(\xi)\label{eq:ar2}
\end{equation}

\begin{equation}
\tau^{\delta_{*}}\Phi_{*}=\tau^{\gamma_{*}+1/2}\int_{-1}^{1}\Gamma_{*}(\zeta)\mathrm{d}\zeta\label{eq:ar6}
\end{equation}

\begin{equation}
\tau^{\alpha_{*}}\Pi_{*}=-\frac{1}{2\pi}\left[\ln2+(1-\tau^{\delta_{*}}\Phi_{*})\ln|\xi|+\tau^{\gamma_{*}+1/2}\int_{-1}^{1}\ln|\xi-\zeta|\,\Gamma_{*}(\zeta)\mathrm{d}\zeta\right]\label{eq:ar3}
\end{equation}
\begin{equation}
\tau^{\alpha_{*}}\Pi_{*}=-\tau^{\beta_{*}-1/2}\frac{1}{4\pi}\int_{-1}^{1}\frac{\Omega_{*}(\zeta)\mathrm{d}\zeta}{\left(\xi-\zeta\right)^{2}}\label{eq:ar4}
\end{equation}

We examine now how time balances in those equations. An examination
of Reynolds equation \eqref{eq:ar2}, balance equation \eqref{eq:ar6},
and elasticity equation \eqref{eq:ar4} reveals that these equations
are homogeneous of degree zero in time --- a requirement for the
existence of a similarity solution of the form \eqref{eq:ar1}, provided
that

\begin{equation}
\delta_{*}=\gamma_{*}+1/2\label{eq:ar11}
\end{equation}
\begin{equation}
\alpha_{*}+3\beta_{*}-\gamma_{*}-1=0\label{eq:ar7}
\end{equation}
\begin{equation}
\alpha_{*}-\beta_{*}+\tfrac{1}{2}=0\label{eq:ar8}
\end{equation}

In contrast, the porous media flow \eqref{eq:ar3} contains three
terms, each with a different power law dependence on time. In view
of the constraints \eqref{eq:ar11}-\eqref{eq:ar8}, it is not possible
to strictly enforce homogeneity of degree zero in time in \eqref{eq:ar3}.
However, it is conceivable that one term vanishes in the limit of
$\tau\rightarrow0$ and $\tau\rightarrow\infty$, with the two other
terms balancing each other and the resulting balance being then homogeneous
of degree zero in time. 
\begin{itemize}
\item In the rock flow regime, $\gamma_*>0$ since $\lim_{\tau\rightarrow0}\Gamma=0$.
Thus $\tau^{\alpha_{*}}\Pi$ balances $\ln|\xi|$ at small time. We
conclude therefore that 
\begin{equation}
\alpha_{r}=0,\quad\beta_{r}=\frac{1}{2},\quad\gamma_{r}=\frac{1}{2},\quad\delta_{r}=1.\label{eq:ar9}
\end{equation}
\item In the fracture flow regime, $\alpha_*<0$ since $\lim_{\tau\rightarrow\infty}\Pi=0$.
The right-hand side of \eqref{eq:ar3} must therefore balance in time.
It then follows that 
\begin{equation}
\alpha_{f}=-\frac{1}{4},\quad\beta_{f}=\frac{1}{4},\quad\gamma_{f}=-\frac{1}{2},\quad\delta_{r}=0.\label{eq:ar10}
\end{equation}
 
\end{itemize}
In the above, the subscript $r$ denotes the small time solution (rock-flow
regime) and the subscript $f$ the large time solution (fracture-flow
regime).

\subsection{$\mathcal{R}$-Regime (Small-Time Solution)\label{subsec:R-Regime}}

The above analysis shows therefore the existence of a similarity solution
at small time. This solution, denoted as the $\mathcal{R}$-vertex
solution, is of the form 
\begin{equation}
\Omega=\Omega_{r}(\xi)\tau^{1/2},\qquad\Pi=\Pi_{r}(\xi),\qquad\Gamma=\Gamma_{r}(\xi)\tau^{1/2},\qquad\Phi=\Phi_{r}\tau\label{eq:rs7}
\end{equation}
Replacing $\Omega$, $\Pi$, and $\Gamma$ by their asymptotic expressions
\eqref{eq:rs7} in the system of equations \eqref{eq:sm1}-\eqref{eq:sm4},
and furthermore noting that the source term in the Reynolds and in
the porous medium flow equations vanishes at small time ---as it
proportional to $\tau$--- leads to the following time-independent
equations 
\begin{equation}
\frac{\partial}{\partial\xi}\left(\Omega_{r}^{3}\frac{\partial\Pi_{r}}{\partial\xi}\right)=\Gamma_{r}-\Phi_{r}\delta(\xi)\label{eq:rs8}
\end{equation}

\begin{equation}
\Phi_{r}=\int_{-1}^{1}\Gamma_{r}(\zeta)\mathrm{d}\zeta\label{eq:rs15}
\end{equation}

\begin{equation}
\Pi_{r}(\xi)=-\frac{1}{2\pi}\ln(2|\xi|)\label{eq:rs10}
\end{equation}
\begin{equation}
\Pi_{r}(\xi)=-\frac{1}{4\pi}\int_{-1}^{1}\frac{\Omega_{r}(\zeta)\mathrm{d}\zeta}{(\xi-\zeta)^{2}}\label{eq:rs11}
\end{equation}
With the additional condition of zero flux that at the crack tip 
\begin{equation}
\Omega_{r}^{3}\frac{\partial\Pi_{r}}{\partial\xi}=0,\quad\xi=\pm1\label{eq:rs14}
\end{equation}
the system of equations \eqref{eq:rs8}-\eqref{eq:rs14} is closed
and can be solved for $\mathcal{S}_{r}(\xi)=\left\{ \Omega_{r},\Pi_{r},\Gamma_{r},\Phi_{r}\right\} $.
However, the crack aperture is determined by integrating the inverse
integral equation \citep{SnLo69}

\begin{equation}
\Omega_{r}(\xi)=\frac{4}{\pi}\int_{0}^{1}\Pi_{r}(\zeta)\ln\left|\frac{\sqrt{1-\xi^{2}}+\sqrt{1-\zeta^{2}}}{\sqrt{1-\xi^{2}}-\sqrt{1-\zeta^{2}}}\right|\,\mathrm{d}\zeta\label{eq:rr1}
\end{equation}
rather than by solving the hypersingular integral eqaution \eqref{eq:rs11}.
Reynolds lubrication equation \eqref{eq:sm1} with $\Gamma=0$ can
only be strictly satisfied at $\tau=0$ and asymptotically at small
time $\tau$.

\begin{figure}[p]
\begin{centering}
\includegraphics{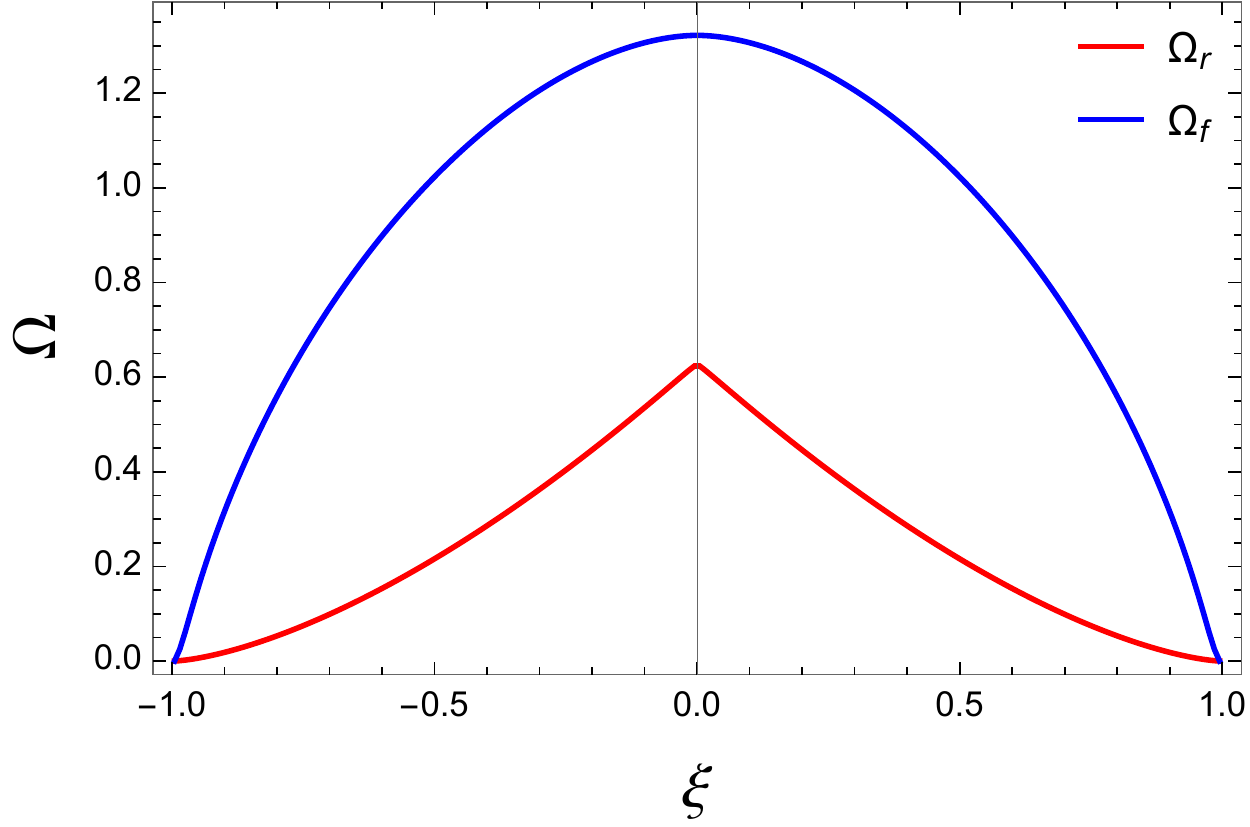}
\par\end{centering}
\caption{Crack opening profile at the $\mathcal{R}$- and $\mathcal{F}$-vertex
(small- and large-time asymptotes). \label{fig:ZT-Asymptotic-aperture-profile}}
\end{figure}

\begin{figure}[p]
\begin{centering}
\includegraphics{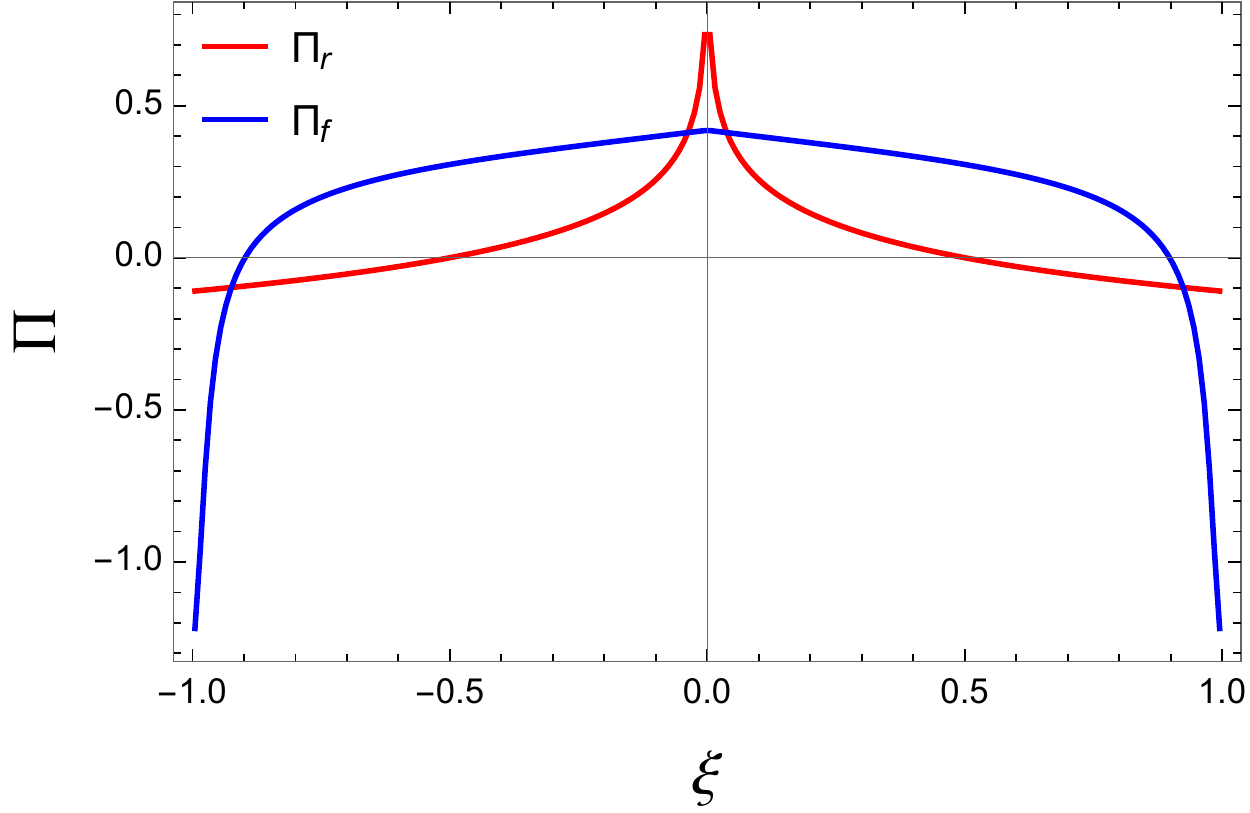}
\par\end{centering}
\caption{Net pressure profile at the $\mathcal{R}$- and $\mathcal{F}$-vertex
(small- and large-time asymptotes). \label{fig:ZT-Asymptotic-net-pressure-profile}}
\end{figure}

In summary, this similarity solution is given by

\begin{equation}
\Pi_{r}=-\frac{1}{2\pi}\ln2|\xi|,\qquad\Omega_{r}=\sqrt{\frac{\pi}{8}}\left[\sqrt{1-\xi^{2}}-\frac{\pi}{2}|\xi|+|\xi|\thinspace\arctan\left(\frac{|\xi|}{\sqrt{1-\xi^{2}}}\right)\right]\label{eq:rr2}
\end{equation}
The small-time asymptotic solution $\Omega_{r}(\xi)$ and $\Pi_{r}(\xi)$
is illustrated in Figs \ref{fig:ZT-Asymptotic-aperture-profile} and
\ref{fig:ZT-Asymptotic-net-pressure-profile}.

The asymptotic strength of the double dipole that reflect the pore
pressure perturbation caused by leak-off from the fracture is given
by $\mathcal{M}^{(2)}=\tau^{1/2}\mathcal{M}_{r}^{(2)}$ with 
\begin{equation}
\mathcal{M}_{r}^{(2)}=\int_{-1}^{1}\xi^{2}\Gamma_{r}(\xi)\,\mathrm{d}\xi\label{eq:ip4}
\end{equation}

\subsection{$\mathcal{F}$-Regime (Large-Time Solution)}

The large-time solution is a similarity solution of the form 
\begin{equation}
\Omega=\Omega_{f}(\xi)\tau^{1/4},\quad\Pi=\Pi_{f}(\xi)\tau^{-1/4},\quad\Gamma=\Gamma_{f}\tau^{-1/2},\quad\Phi=\Phi_{f}.\label{eq:fs2}
\end{equation}
However, necessarily $\Phi_{f}=1$, since all the injected fluid enters
the crack directly. The equations governing the solution $\mathcal{S}_{f}(\xi;\mathcal{K})=\{\Omega_{f},\,\Pi_{f},\,\Gamma_{f},\,\Phi_{f}\}$
are 
\begin{equation}
\frac{\mathrm{d}}{\mathrm{d}\xi}\left(\Omega_{f}^{3}\frac{\mathrm{d}\Pi_{f}}{\mathrm{d}\xi}\right)=\Gamma_{f}-\delta(\xi)\label{eq:fs4}
\end{equation}

\begin{equation}
\int\ln|\xi-\zeta|\,\Gamma_{f}(\zeta)\mathrm{d}\zeta=-\ln2\label{eq:fs5}
\end{equation}
\begin{equation}
\Pi_{f}(\xi)=-\frac{1}{4\pi}\int_{-1}^{1}\frac{\Omega_{f}(\zeta)\mathrm{d}\zeta}{(\xi-\zeta)^{2}}\label{eq:fs6}
\end{equation}
With the additional conditions that at the crack tip 
\begin{equation}
\Omega_{f}^{3}\frac{\mathrm{d}\Pi_{f}}{\mathrm{d}\xi}=0,\quad\xi=\pm1\label{eq:fs8}
\end{equation}
The system of equations \eqref{eq:fs4}-\eqref{eq:fs8} is closed
and can be solved for $\mathcal{S}_{f}(\xi)$. 

The porous media flow equation \eqref{eq:fs5} can be solved in closed-form
for $\Gamma_{f}$ 
\begin{equation}
\Gamma_{f}=\frac{1}{\pi\sqrt{1-\xi^{2}}}\label{eq:fs9}
\end{equation}
Then after defining the flux $\Psi_{f}$ (and noting also that $\Psi=\Psi_{f}$)
as 
\begin{equation}
\Psi_{f}=-\Omega_{f}^{3}\frac{\mathrm{d}\Pi_{f}}{\mathrm{d}\xi}\label{eq:fs10}
\end{equation}
integration of \eqref{eq:fs4} yields 
\begin{equation}
\Psi_{f}=\mathrm{sgn}(\xi)\left(\frac{1}{2}-\frac{1}{\pi}\arcsin|\xi|\right)\label{eq:fs11}
\end{equation}
The two fields $\Omega_{f}(\xi)$ and $\Pi_{f}(\xi)$ are then computed
by solving numerically the elasticity equation \eqref{eq:fs6} and
the integrated Reynolds equation 
\begin{equation}
\Omega_{f}^{3}\frac{\mathrm{d}\Pi_{f}}{\mathrm{d}\xi}=-\mathrm{sgn}(\xi)\left(\frac{1}{2}-\frac{1}{\pi}\arcsin|\xi|\right)\label{eq:fs12}
\end{equation}
Appendix \ref{sec:Numerical-Scheme} outlines the numerical scheme.
Figures \ref{fig:ZT-Asymptotic-aperture-profile}-\ref{fig:ZT-Asymptotic-net-pressure-profile}
illustrate the large-time asymptotic profile for the opening $\Omega_{\mathrm{f}}(\xi)$
and the net pressure $\Pi_{\mathrm{f}}(\xi)$, respectively.

Finally, $\mathcal{M}^{(2)}=\tau^{-1/2}\mathcal{M}_{f}^{(2)}$ with
\begin{equation}
\mathcal{M}_{f}^{(2)}=\int_{-1}^{1}\xi^{2}\Gamma_{f}(\xi)\,\mathrm{d}\xi\label{eq:ip5}
\end{equation}

\section{Numerical Algorithm}

\subsection{Preamble }

The numerical scheme is based on the mathematical model without $\Phi$
as a primary variable, as formuated in Section \eqref{subsec:Alternative-Scaled-Mathematical}
Discretization of the set of integro-differential equations \eqref{eq:sm12}-\eqref{eq:sm14}
and boundary conditions \eqref{eq:sm19} leads to a nonlinear algebraic
system of equations in terms of the crack opening $\Omega$ at discrete
points on the crack. The starting point is to segment the crack on
$[-1,1]$ into $2n$ elements, each of equal length with the mid-point
of element $i$ located at $\xi_{i}=(2i-1)h-1$. The discretization
of the system of equations rests on assuming that both $\tilde{\Gamma}$
---the discontinuity of the normal flux, and $\Omega$ ---the discontinuity
of the displacement normal to the crack, are uniform along each element.
Thus three quantities are assigned to element $i$: $\tilde{\Gamma}_{i}$,
$\Pi_{i}$, and $\Omega_{i}$, with $\Pi_{i}$ representing the net
pressure at $\xi_{i}$.

Introducing the piecewise constant approximation of the discontinuity
fields $\Gamma(\xi)$ and $\Omega(\xi)$ into the singular integral
equations \eqref{eq:sm13} and \eqref{eq:sm14} leads to two linear
systems of equations 
\begin{equation}
\boldsymbol{\Pi}-\boldsymbol{\Pi}_{o}=\tau^{1/2}\boldsymbol{B}\cdot\tilde{\boldsymbol{\Gamma}},\qquad\tau^{1/2}\boldsymbol{\Pi}=\boldsymbol{A}\cdot\boldsymbol{\Omega}\label{eq:na1}
\end{equation}
where $\boldsymbol{\Pi}=\{\Pi_{1},\cdots,\Pi_{2n}\}^{T}$, $\tilde{\boldsymbol{\Gamma}}=\{\tilde{\Gamma}_{1},\cdots,\tilde{\Gamma}_{2n}\}^{T}$,
and $\boldsymbol{\Omega}=\{\Omega_{1},\cdots,\Omega_{2n}\}^{T}$,
with vector $\boldsymbol{\Pi}_{o}$ consisting of values of function
$\Pi_{o}(\xi)$ evaluated at the element mid-points located at $\xi_{i}$,
$i=1,\cdots,2n$. Also, $\boldsymbol{A}$ denotes the elasticity matrix
linking the net pressure to the crack opening and $\boldsymbol{B}$
is the porous flow matrix linking the fluid pressure to the leak-off.
These two equations are then combined with a discretized Reynolds
equation \eqref{eq:sm12} to yield a nonlinear system of equations
to be solved for $\boldsymbol{\Omega}$.

\subsection{Elasticity}

As a consequence of assuming that $\Omega(\xi)$ is piecewise constant,
integral equation \eqref{eq:sm14} simplifies to 
\begin{equation}
\Pi(\xi)=-\frac{1}{4\pi\tau^{1/2}}\sum_{j=1}^{2n}\Omega_{j}\int_{\xi_{j}-h}^{\xi_{j}+h}\frac{\mathrm{d}\zeta}{(\xi-\zeta)^{2}}\label{eq:na2}
\end{equation}
or after integration to 
\begin{equation}
\Pi(\xi)=-\frac{1}{4\pi\tau^{1/2}}\sum_{j=1}^{2n}\Omega_{j}\left(\frac{1}{\xi-\xi_{j}-h}-\frac{1}{\xi-\xi_{j}+h}\right)\label{eq:na2-1}
\end{equation}
Hence, evaluating $\Pi_{i}=\Pi(\xi_{i})$ from \eqref{eq:na2-1} yields
\begin{equation}
\tau^{1/2}\Pi_{i}=\sum_{j=1}^{2n}A_{ij}\Omega_{j},\quad i=1,\cdots,2n\label{eq:na3}
\end{equation}
where 
\begin{equation}
A_{ij}=-\frac{1}{2\pi}\frac{h}{(\xi_{i}-\xi_{j})^{2}-h^{2}}\label{eq:na4}
\end{equation}

To improve accuracy of the solution, the self-coefficient of the two
edge elements ($i=1$ and $i=2n$) is increased by $1/24\,h$ \citep{GoDe11b},
following the theoretical argument put forward by \citet{RyNa85}.

\subsection{Porous Medium Flow}

With $\Gamma(\xi)$ also assumed to be piecewise constant, the integral
equation \eqref{eq:sm13} simplifies to

\begin{equation}
\Pi(\xi)-\Pi_{o}(\xi)=-\frac{\tau^{1/2}}{2\pi}\sum_{1=1}^{2n}\tilde{\Gamma}_{j}\int_{\xi_{j}-h}^{\xi_{j}+h}\ln(|\xi-\zeta|)\mathrm{d}\zeta\label{eq:na5}
\end{equation}
Hence, evaluating $\Pi_{i}=\Pi(\xi_{i})$ from \eqref{eq:na5} and
$\Pi_{oi}=\Pi_{o}(\xi_{i})$ from \eqref{eq:sm19} yields 
\begin{equation}
\Pi_{i}-\Pi_{oi}=\tau^{1/2}\sum_{1=1}^{2n}B_{ij}\tilde{\Gamma}_{j},\quad i=1,\cdots,2n\label{eq:na6}
\end{equation}
where 
\begin{equation}
B_{ij}=\frac{1}{2\pi}\left[2h+\left(\xi_{i}-\xi_{j}-h\right)\ln|\xi_{i}-\xi_{j}-h|-\left(\xi_{i}-\xi_{j}+h\right)\ln|\xi_{i}-\xi_{j}+h|\right].\label{eq:na7}
\end{equation}

\subsection{Lubrication }

Discretizing Reynolds equation \eqref{eq:sm12} using a finite volume
approach leads to

\begin{equation}
\boldsymbol{C}\cdot\boldsymbol{\Pi}=\tau\tilde{\boldsymbol{\Gamma}},\label{eq:na8}
\end{equation}
where the matrix $\boldsymbol{C}(\boldsymbol{\Omega})$ is given by

\begin{equation}
\boldsymbol{C}=\frac{1}{4h^{2}}\left[\begin{array}{ccccc}
-K_{3/2} & K_{3/2}\\
K_{3/2} & -(K_{3/2}+K_{5/2}) & K_{5/2}\\
 & \ddots & \ddots & \ddots\\
 &  & K_{2n-3/2} & -(K_{2n-3/2}+K_{2n-1/2}) & K_{2n-1/2}\\
 &  &  & K_{2n-1/2} & -K_{2n-1/2}
\end{array}\right]\label{eq:na9}
\end{equation}
with 
\begin{equation}
K_{i-1/2}=(\Omega_{i-1}^{3}+\Omega_{i}^{3})/2.\label{eq:na10}
\end{equation}

\subsection{Discrete System of Equations}

In summary, the discrete solution at time $\tau$ is governed by the
following system of algebraic equations 
\begin{align}
 & \boldsymbol{C}(\boldsymbol{\Omega})\cdot\boldsymbol{\Pi}=\tau\tilde{\boldsymbol{\Gamma}}\label{eq:na15}\\
 & \boldsymbol{\Pi}-\boldsymbol{\Pi}_{o}=\tau^{1/2}\boldsymbol{B}\cdot\tilde{\boldsymbol{\Gamma}}\label{eq:na16}\\
 & \tau^{1/2}\boldsymbol{\Pi}=\boldsymbol{A}\cdot\boldsymbol{\Omega}\label{eq:na17}
\end{align}
Combining \eqref{eq:na15}-\eqref{eq:na17} to eliminate $\boldsymbol{\Pi}$
and $\tilde{\boldsymbol{\Gamma}}$ results in a system of equations
in terms of $\boldsymbol{\Omega}$ only 
\begin{equation}
\left(\boldsymbol{I}-\tau^{-1/2}\boldsymbol{B}\cdot\boldsymbol{C}(\boldsymbol{\Omega})\right)\cdot\boldsymbol{A}\cdot\boldsymbol{\Omega}=\tau^{1/2}\boldsymbol{\Pi}_{o},\label{eq:na19}
\end{equation}
where $\boldsymbol{I}$ denotes the identity matrix of size $2n$.
The system of $2n$ nonlinear algebraic equations \eqref{eq:na19}
can be solved for $\boldsymbol{\Omega}$, using a fixed-point algorithm.
Arrays $\boldsymbol{\Pi}$ and $\tilde{\boldsymbol{\Gamma}}$ are
then readily deduced from $\boldsymbol{\Omega}$ using \eqref{eq:na15}
and \eqref{eq:na17}.

\section{Results}

\begin{figure}[p]
\begin{centering}
\includegraphics{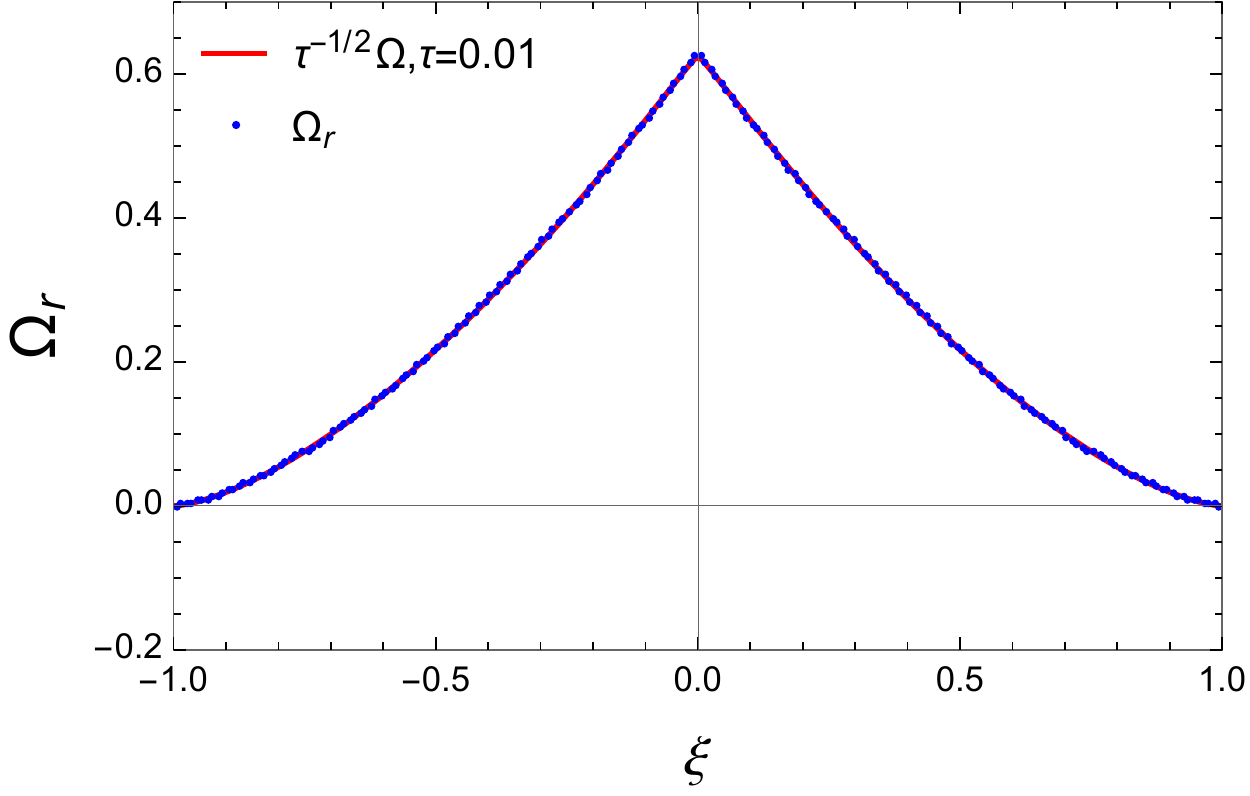} 
\par\end{centering}
\caption{Comparison between $\mathcal{R}$-vertex opening $\Omega_{r}(\xi)$
computed either directly or deduced from the transient calculation
at time $\tau=10^{-2}$. \label{fig:R-Opening}}
\end{figure}

\begin{figure}[p]
\begin{centering}
\includegraphics{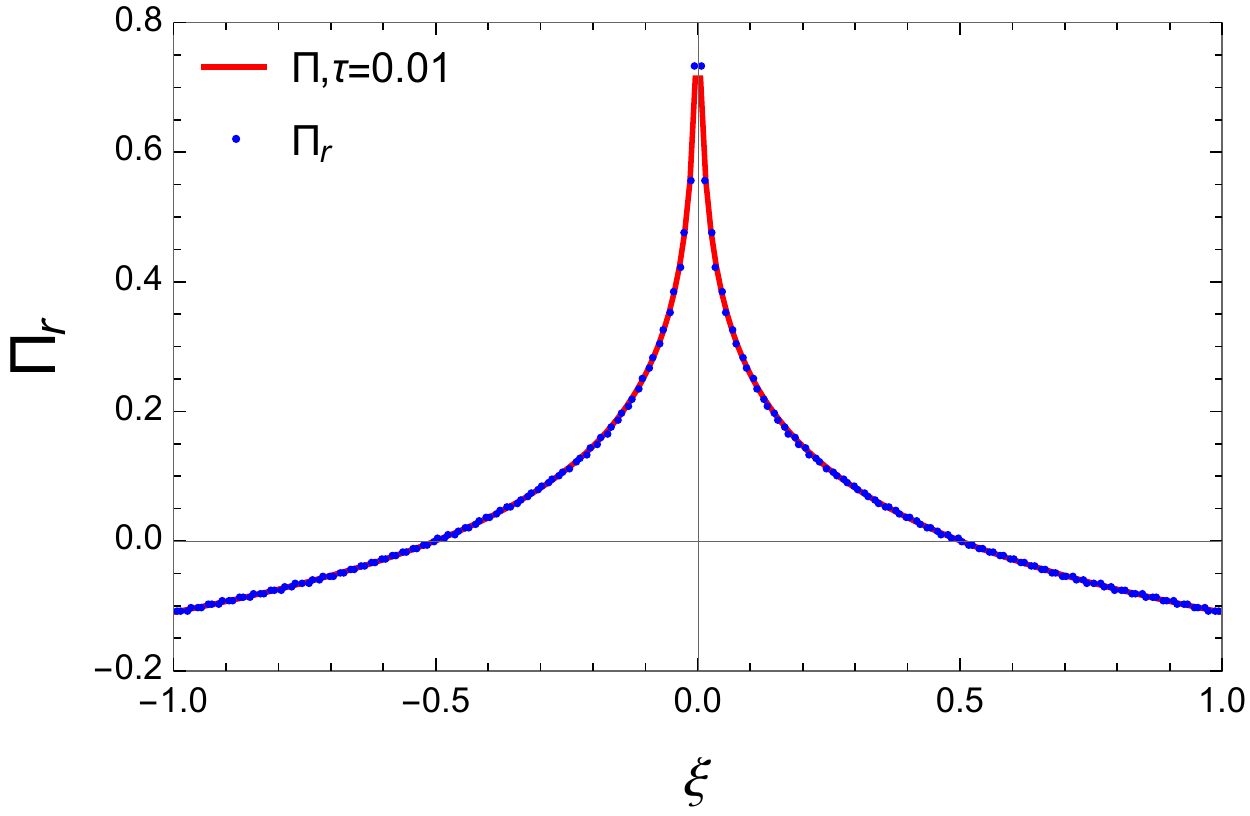} 
\par\end{centering}
\caption{Comparison between $\mathcal{R}$-vertex net pressure $\Pi_{r}(\xi)$
computed either directly or deduced from the transient calculation
at time $\tau=10^{-2}$. \label{fig:R-Pressure}}
\end{figure}

Figures \ref{fig:R-Opening} and \ref{fig:R-Pressure} show the profiles
of aperture and net pressure in the $\mathcal{R}$-regime and compare
the analytical solutions for $\Omega_{r}(\xi)$ and $\Pi_{r}(\xi)$
given by \eqref{eq:rr2} with the transient solution computed numerically
at $\tau=10^{-2}$. The numerical solution was calculated using 200
source and displacement discontinuity elements. Since $\Omega\sim\text{\ensuremath{\tau}}^{1/2}$
at small time, $\Omega$ is multiplied by $10^{-1}$ to enable comparison
between the numerical and analytical solutions. In contrast, $\Pi$
does not depend on time at small $\tau$. These figures show the high
quality of the numerical solutions and confirm that the solution is
still in the $\mathcal{R}$-regime at $\tau=10^{-2}.$

\begin{figure}[p]
\begin{centering}
\includegraphics{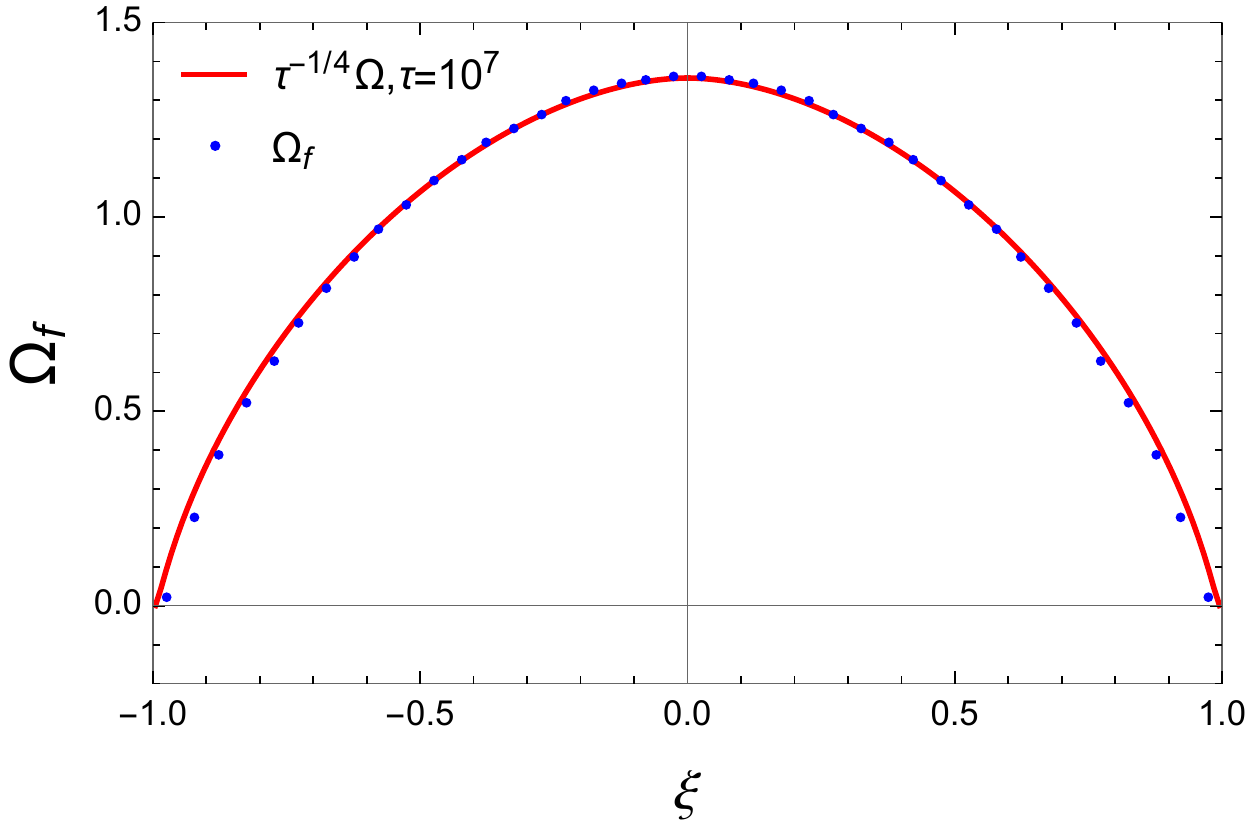} 
\par\end{centering}
\caption{Comparison between $\mathcal{F}$-vertex opening $\Omega_{f}(\xi)$
computed either directly or deduced from the transient calculation
at time $\tau=10^{7}$. \label{fig:F-Opening}}
\end{figure}

\begin{figure}[p]
\begin{centering}
\includegraphics{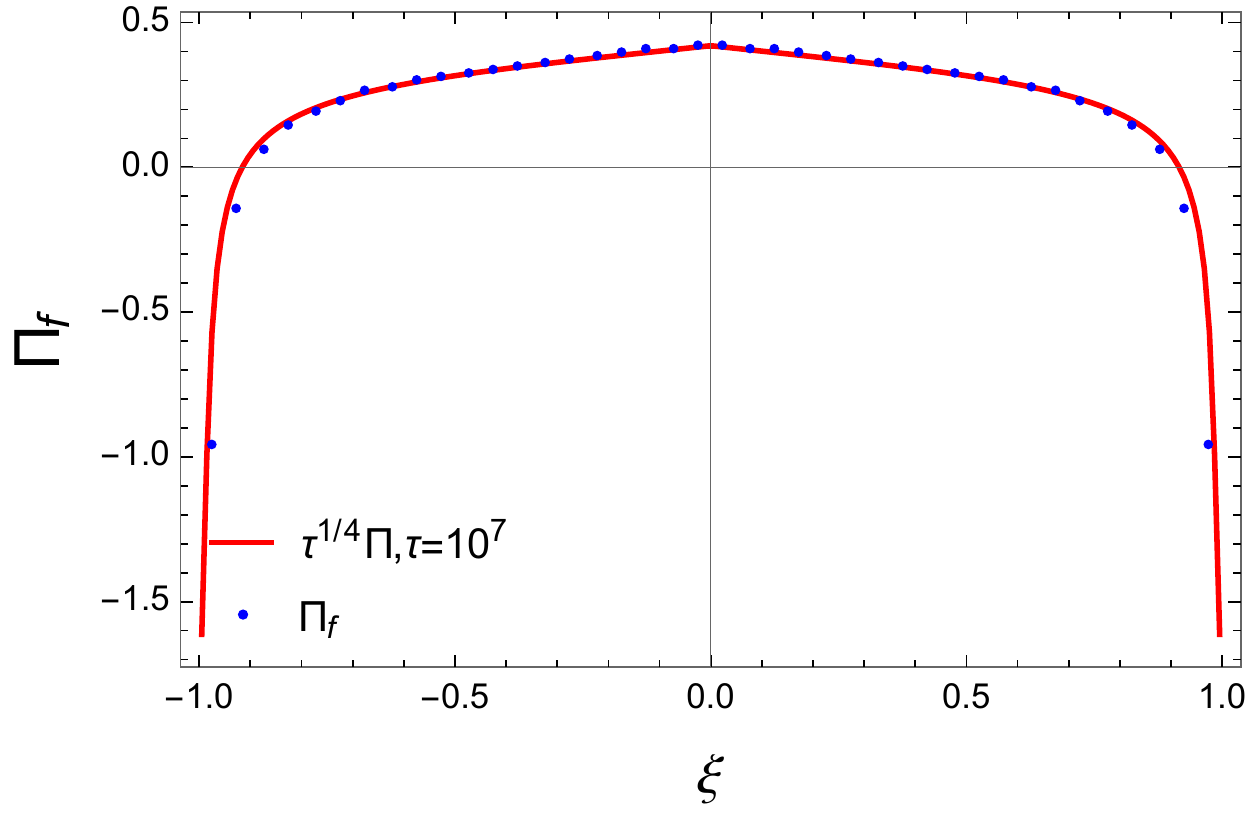} 
\par\end{centering}
\caption{Comparison between $\mathcal{F}$-vertex net pressure $\Pi_{f}(\xi)$
computed either directly or deduced from the transient calculation
at time $\tau=10^{7}$. \label{fig:F-Pressure}}
\end{figure}

Figures \ref{fig:F-Opening} and \ref{fig:F-Pressure} provide a similar
comparison between the $\mathcal{F}$-vertex solution (computed numerically
using a dedicated algorithm) and the transient solution computed at
time $\tau=10^{7}$. The transient solution was calculated using 200
source and displacement discontinuity elements while the $\mathcal{F}$-vertex
solution was obtained using 40 elements. Now, $\Omega$ is multiplied
by $\tau^{-1/4}$ and $\Pi$ by $\tau^{1/4}$ to enable comparison
between the two solutions. These figures show again the high quality
of the numerical solution and confirm that the solution is in the
$\mathcal{F}$-regime at $\tau=10^{7}.$

\begin{figure}[p]
\begin{centering}
\includegraphics{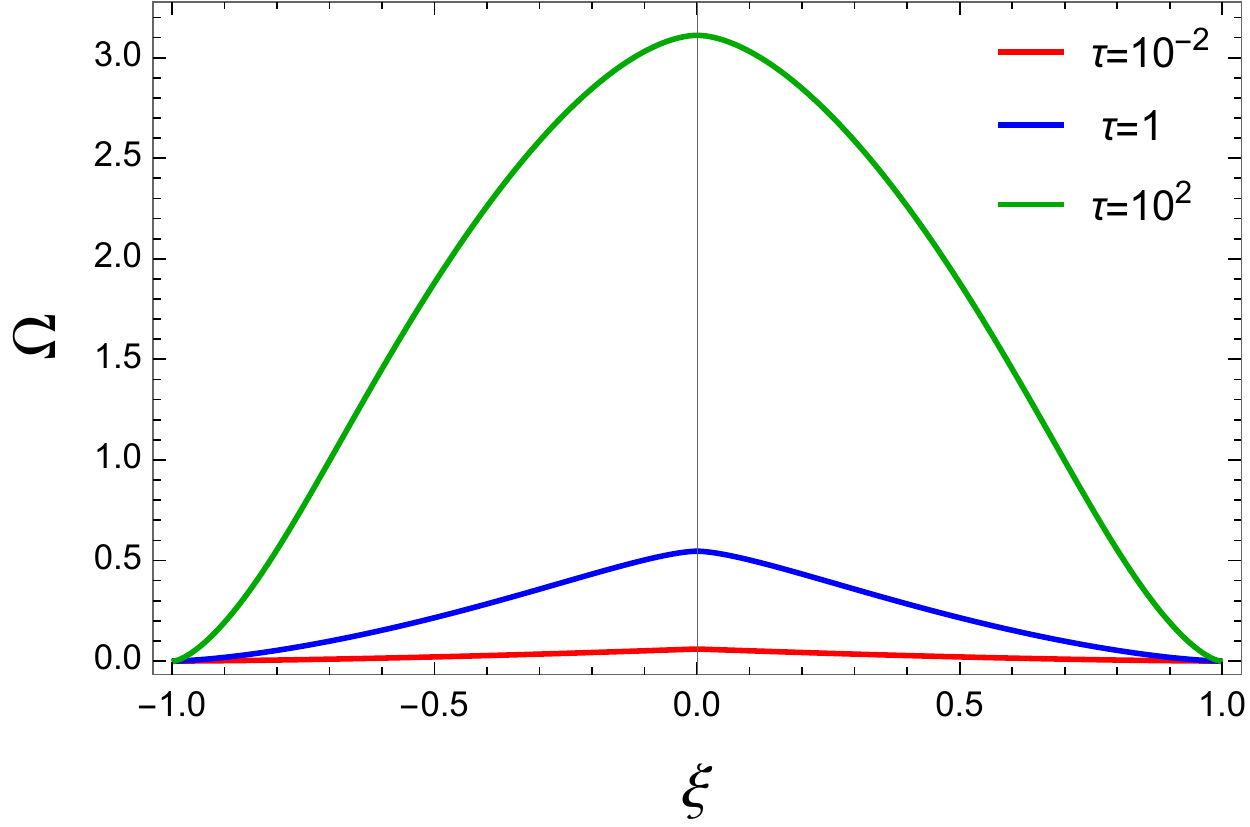} 
\par\end{centering}
\caption{Opening profile in time scaling at $\tau=10^{-2},\,1,\,10^{2}$. \label{fig:ZT-time-Opening-profile}}
\end{figure}

\begin{figure}[p]
\begin{centering}
\includegraphics{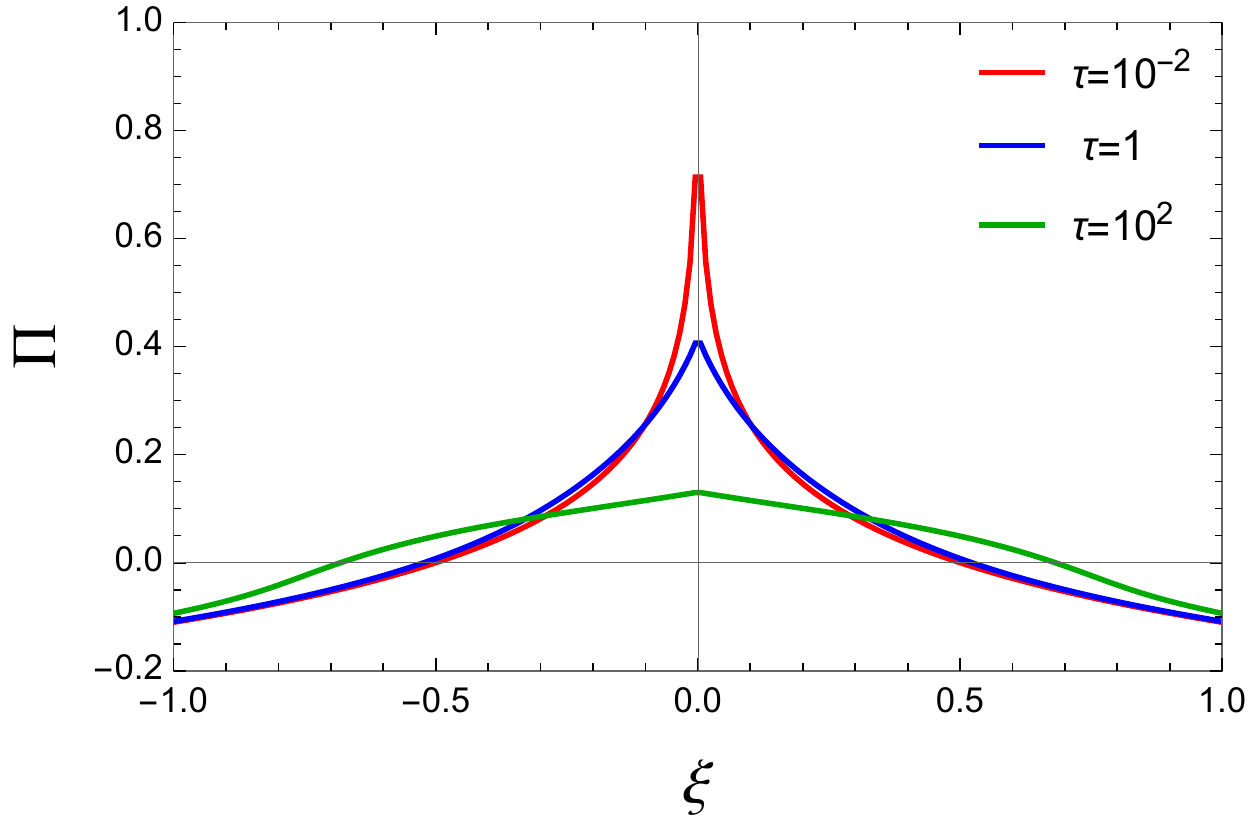} 
\par\end{centering}
\caption{Net pressure profile in time scaling at $\tau=10^{-2},\,1,\,10^{2}$.
\label{fig:ZT-time-Net-pressure-profile}}
\end{figure}

The changing nature of the solution with time is illustrated in Figs.
\ref{fig:ZT-time-Opening-profile} and \ref{fig:ZT-time-Net-pressure-profile},
which show profiles of $\Omega(\xi)$ and $\Pi(\xi)$ at time $\tau=10^{-2},\,1,\,10^{2}$
in a time-independent scaling. We can observe a large spatial variation
of the net pressure at $\tau=10^{-2}$ compared to $\Pi(\xi)$ at
$\tau=10^{2}$, but a small aperture at $\tau=10^{2}$ compared to
$\Omega(\xi)$ at $\tau=10^{2}$. 

%
%
%

\section{Conclusions}

This paper has described a KGD-type model to simulate for the propagation
of a hydraulic fracture in very permeable rocks. The model is based
on the key assumptions that the volume of fluid stored in the hydraulic
fracture is negligibly small compared to the injected volume and that
the crack is propagating within a domain where the hydraulic fields
are quasi-stationary. Scaling of the equations indicates that the
solution depends only on a dimensionless time $\tau$, as all the
physical parameters defining the problem are absorbed in the scales.
A peculiarity of this problem is the extreme sensitivity of the timescale
on a dimensionless injection rate.

Two asymptotic regimes bound the solution: the rock-flow dominated
regime or $\mathcal{R}$-regime at small time (approximately $\tau\lesssim10^{-2}$),
and the fracture-flow dominated regime or $\mathcal{F}$-regime at
large time (approximately $\tau\apprge10^{2}$). The fracture has
a zero conductivity and is thus hydraulically invisible in the $\mathcal{R}$-regime,
while it has a large conductivity in the $\mathcal{F}$-regime. Thus
all the injected fluid enters the reservoir via the borehole at small
time, but via the crack at large time. Furthermore, the injection
pressure increases as $\ln\text{\ensuremath{\tau}}$ in the $\mathcal{R}$-regime,
but decreases as $\tau^{-1/4}$ in the $\mathcal{F}$-regime. The
peak injection pressure, which should not be interpreted as the breakdown
pressure, takes place in the transition between the two regimes.

This study suggests that the hydraulic aspects are important, likely
critical, in this class of problems, which have so far been viewed
only through the prism of strength.

\bibliographystyle{plainnat}

\newpage
\appendix

\section{Numerical Scheme for $\mathcal{F}$-Vertex Solution\label{sec:Numerical-Scheme}}

The two fields $\Omega_{f}(\xi)$ and $\Pi_{f}(\xi)$ are computed
by solving numerically the elasticity equation \eqref{eq:fs6}, the
integrated Reynolds equation \eqref{eq:fs12}, and the propagation
criterion \eqref{eq:sm4} expressed in terms $\Pi_{f}(\xi)$. Discretization
of \eqref{eq:fs6}, \eqref{eq:fs12} and \eqref{eq:sm4} leads to
system of algebraic equations consists of $2n$ linear equations of
elasticity

\begin{equation}
\boldsymbol{\Pi}=\boldsymbol{A}\cdot\boldsymbol{\Omega}\label{eq:A1}
\end{equation}
$2n-1$ non-linear equations formulated by discretizing the integrated
Reynolds equation at the mid-nodes, and 1 linear equation expressing
the propagation criterion. The combined discretized non-linear equations
and propagation criterion can be written as

\begin{equation}
\boldsymbol{D}(\boldsymbol{\Omega})\cdot\boldsymbol{\Pi}=-\boldsymbol{\Psi},\label{eq:A2}
\end{equation}
where the matrix $\boldsymbol{D}(\boldsymbol{\Omega})$ is given by

\begin{equation}
\boldsymbol{D}=\frac{1}{2h}\left[\begin{array}{ccccc}
-K_{3/2} & K_{3/2}\\
 & -K_{5/2} & K_{5/2}\\
 &  & \ddots & \ddots\\
 &  &  & -K_{2n-1/2} & K_{2n-1/2}\\
W_{1} & W_{2} & \cdots & W_{2n-1} & W_{2n}
\end{array}\right],\qquad\boldsymbol{\Psi}=\left[\begin{array}{c}
\Psi_{f,1/2}\\
\Psi_{f,3/2}\\
\vdots\\
\Psi_{f,2n-1/2}\\
0
\end{array}\right]\label{eq:A3}
\end{equation}
with 
\begin{equation}
K_{i-1/2}=(\Omega_{f,i-1}^{3}+\Omega_{f,i}^{3})/2\label{eq:A4}
\end{equation}
and 
\begin{equation}
W_{i}=\arcsin(\xi_{i}+a)-\arcsin(\xi_{i}-a)\label{eq:A5}
\end{equation}
Note that the flux at the common boundary of elments $n$ and $n+1$
, $\Psi_{f,n+1/2}=0$, as it is the average of a discontinuous flux
of opposite sign on both sides of the origin.

Combining \eqref{eq:A1} and \eqref{eq:A2} leads to a system of $2n$
non-linear algebraic equations 
\[
\boldsymbol{D}(\boldsymbol{\Omega})\cdot\boldsymbol{A}\cdot\boldsymbol{\Omega}=-\boldsymbol{\Psi}
\]
to be solved for $\boldsymbol{\Omega}$.
\end{document}